\documentclass[prd,twocolumn,amssymb,amsmath]{revtex4-1}
\usepackage{graphicx}
\usepackage{epsfig,color}
\usepackage{amsmath,amssymb,eufrak}
\usepackage{amsthm}
\usepackage{enumerate}

\begin{document}

\title{Combining Tensor Networks with Monte Carlo Methods for Lattice Gauge Theories}

\date{\today}

\author{Erez Zohar}
\address{Max-Planck-Institut f\"ur Quantenoptik, Hans-Kopfermann-Stra\ss e 1, 85748 Garching, Germany.}

\author{J. Ignacio Cirac}
\address{Max-Planck-Institut f\"ur Quantenoptik, Hans-Kopfermann-Stra\ss e 1, 85748 Garching, Germany.}

\begin{abstract}
Gauged gaussian Projected Entangled Pair States are particular tensor network constructions that describe lattice states of fermionic matter interacting with dynamical gauge fields.  We show how one can efficiently compute, using Monte-Carlo techniques, expectation values of physical observables in that class of states. This opens up the possibility of using tensor network techniques to investigate lattice gauge theories in two and three spatial dimensions.
\end{abstract}

\maketitle

Tensor networks, in particular Matrix Product States (MPS) \cite{Fannes1992} and Projected Entangled Pair States (PEPS)  \cite{PEPS2004}, are special many-body states, which have recently been applied mostly for
condensed matter physics. Such states, constructed by contracting local tensors that depend on few parameters, may well represent relevant physical states, e.g. ground states of local Hamiltonians \cite{Cirac2009,Orus2014}.
 This makes them very useful for several purposes, including the characterization of ground \cite{White1992} and thermal equilibrium states \cite{Thermal2004,Zwolak2004} and, to some extent, the dynamics \cite{Daley2004} of many-body quantum systems in one spatial dimension. Recently, they have also been successfully used in several problems in two spatial dimensions (see, e.g., \cite{Corboz2014,Niesen2017}). The extension to three dimensions, and even to some problems in two dimensions, remains very challenging due to the unfavorable scaling of the computational time \cite{PEPS2004}.

Recently, it was recognized that HEP (high energy physics) problems, in particular lattice gauge theories (LGTs) \cite{Wilson,KogutLattice} (with both matter and gauge fields) could also be addressed using tensor networks \cite{Byrnes2002,Banuls2013,Tagliacozzo2014,Haegeman2014,Zohar2015b,Zohar2016a,Kull2017}, in the Hamiltonian formulation \cite{KogutSusskind}. While conventional, Euclidean Monte-Carlo LGT methods have been extremely successful \cite{FLAG2013}, complementary methods could still be useful to
tackle real time dynamics or fermions in scenarios that suffer, in the Euclidean scheme, from the sign problem \cite{Troyer2005}.  Indeed, $1+1$d numerical studies of LGTs, both Abelian and non-Abelian, have been very successful, showing remarkable results \cite{Byrnes2002,Sugihara2005,Banuls2013,Buyens2013,Silvi2014,Rico2013,Saito2014,Kuhn2015,Banuls2015,Banuls2016,Pichler2015,Silvi2016,Milsted2016,Buyens2016,Banuls2016,Verstraete2016,Banuls2017,Banuls2017a,Banuls2017b}. However, as in the condensed matter context, going beyond $1+1$d is, in general, much more demanding, and higher dimensional studies of gauge-invariant PEPS are still mostly analytical, or involve exact calculations for small systems
\cite{Tagliacozzo2014,Haegeman2014,Zohar2015b,Zohar2016,Zohar2016a,Zapp2017}.

An important feature of PEPS is that they allow one to encode symmetries in a very simple way, including local (gauge) ones.
The main idea of locally symmetric PEPS follows what is done in HEP.
One starts out with a tensor network state that involves only matter degrees of freedom, with a global symmetry. Then, as in the minimal coupling procedure \cite{Peskin1995}, the global symmetry is lifted to a local one by introducing
new degrees of freedom (gauge field), in the form of a new tensor \cite{Haegeman2014,Zohar2015b,Zohar2016,Zohar2016a}. In \cite{Zohar2015b,Zohar2016a},  such constructions with $U(1)$ and $SU(2)$ symmetries in $2+1$d were discussed;
There, inspired by the standard recipe of HEP models, the tensor network states corresponding to the matter alone are taken to be gaussian, corresponding to ground states of non-interacting fermionic theories. Once such states are gauged with the above procedure, the resulting states, which we will call \emph{gauged gaussian PEPS (GGPEPS)}, describe interacting LGTs. In \cite{Zohar2015b,Zohar2016a}, the simplest GGPEPS with some extra global symmetries (like rotation) were parametrized and their physical properties were studied for different values of those parameters.
In spite of their simplicity, these states manifested
 very rich physics, including gapped and gapless phases; in the pure gauge case, both confinement and deconfinement phases were reported; in the dynamical fermions case, several screening/non-screening behaviours were found. Some of the results were obtained analytically, but most of them involved exact computations on cylinders - which forced the system to be very small.
 The next natural steps would be to increase the bond dimension, consider larger systems, determine the parameters that minimize the energy of relevant Hamiltonians corresponding to HEP models and, eventually, extend all that to $3+1$d. However, due to the unfavorable scaling of the computational time with the bond dimension in dimensions higher than one \cite{Verstraete2008}, this ambitious program is largely unreachable.

In this paper, we introduce a formalism that allows one to efficiently compute expectation values of gauge-invariant observables with GGPEPS. The key idea is to reexpress these expectation values
such that they could be computed with Monte Carlo methods, using Metropolis-like algorithms
(for other methods combining tensor networks and Monte-Carlo approaches see \cite{Sandvik2007,Schuch2008}). With this method, there is no Monte-Carlo related sign problem, so that one could easily treat larger systems and even consider $3+1$d.

 The method is valid for both Abelian and non-Abelian symmetries, and unlike conventional LGT computations, does not require to integrate over the time dimension, making it possibly very efficient. We shall explain the method, review the GGPEPS construction and illustrate with a proof-of-principle example based on a previously
introduced GGPEPS \cite{Zohar2015b}.

\begin{figure}[t!]
  \centering
  \includegraphics[width=0.45\textwidth]{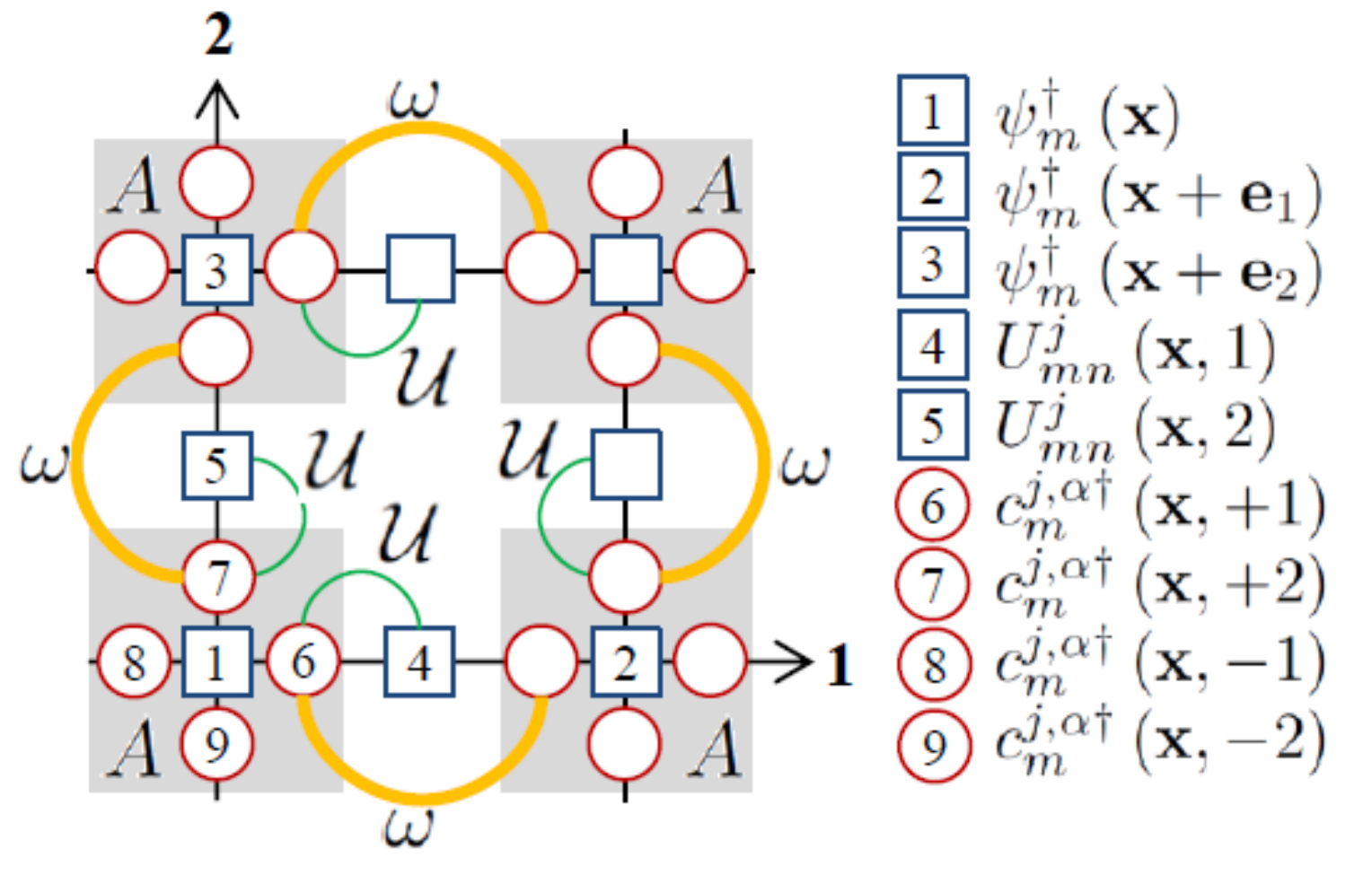}\\
  \caption{A $1-2$ plane plaquette. Squares represent physical degrees of freedom, and circles - virtual. The vertex operators $A$ connect objects within the grey
boxes; the link operators $\omega$ connect fermions attached by thick orange lines; the gauging transformation $\mathcal{U}$ entangles objects connected by thin green lines.
See the legend for labeling conventions.}\label{fig1}
\end{figure}

\emph{The Hilbert space of a lattice gauge theory.} The conventional Hamiltonian setting for LGTs in $d+1$  \cite{KogutSusskind} involves matter and gauge fields.
The Hilbert space is a tensor product of the matter fermionic Fock space on the lattice's vertices $\mathbf{x} \in \mathbb{Z}^d$, with the gauge field Hilbert space which is, itself, a  product of local Hilbert spaces attached
to the links $\left\{\mathbf{x},k\right\}$ ($\mathbf{x}\in\mathbb{Z}^d$ denotes the beginning of a link, and $k=1...d$ - its direction - see Fig. \ref{fig1}).
Let our gauge group
$G=\left\{g\right\}$ be finite or Lie. One can define the complete set of \emph{group element states} \cite{Zohar2015}, $\left\{\left|g\right\rangle\right\}_{g\in G}$ labeled by group elements, spanning the local gauge field Hilbert space on each link $\left\{\mathbf{x},k\right\}$.  In the finite case, this is an orthonormal basis with $\left\langle g' | g\right\rangle = \delta_{g,g'}$; if $G$ is infinite,
 $\left\langle g' | g\right\rangle = \delta\left(g,g'\right)$ - a distribution defined with respect to the group's measure. We further introduce the \emph{configuration states}, $\left|\mathcal{G}\right\rangle \equiv \underset{\mathbf{x},k}{\bigotimes}\left|g\left(\mathbf{x},k\right)\right\rangle$ - configurations of group elements on all the links,
 spanning the global gauge field Hilbert space.

 A state in the whole Hilbert space, with both gauge fields and fermions, may be expanded as
\begin{equation}
\left|\Psi\right\rangle = \int \mathcal{DG}\left|\mathcal{G}\right\rangle \left|\psi\left(\mathcal{G}\right)\right\rangle
\label{psibig}
\end{equation}
where $\mathcal{DG} = \underset{\mathbf{x},k}{\prod}dg\left(\mathbf{x},k\right)$ \footnote{If the group is finite, one has to sum over all group elements instead, here and throughout this paper.}, and the un-normalized state $\left|\psi\left(\mathcal{G}\right)\right\rangle$
represents the fermionic part of the state, depending on $\mathcal{G}$ as parameters.
This still does not guarantee that $\left|\Psi\right\rangle$ is gauge-invariant: this has to be imposed on $\left|\psi\left(\mathcal{G}\right)\right\rangle$.
One way of ensuring that is to follow the PEPS procedure introduced in \cite{Zohar2015b,Zohar2016a}, where, inspired by minimal coupling,
  one starts with non-interacting, gaussian fermionic states and \emph{gauges} them. Our goal is to show that in this construction of $\left|\Psi\right\rangle$, $\left|\psi\left(\mathcal{G}\right)\right\rangle$ is a gaussian state, and make use of that fact to show that  $\left\langle\psi\left(\mathcal{G}\right)|\psi\left(\mathcal{G}\right)\right\rangle$  and other related quantities can be efficiently computed. For  that, however, we must review a few more details on the gauge symmetry.

On the links, we introduce the unitary operators $\Theta_g$ and and $\widetilde{\Theta}_g$, which realize right $\Theta_g\left|h\right\rangle = \left|hg^{-1}\right\rangle$ and left $\widetilde{\Theta}_g\left|h\right\rangle = \left|g^{-1}h\right\rangle$ group transformations, and the trivial state, $\left|000\right\rangle$ (in the \emph{representation basis} - see App. \ref{appa}), which is invariant under any group transformation.

 The matter fields are represented by spinors $\psi^{\dagger}_m\left(\mathbf{x}\right)$, that transform with respect to some representation (e.g. the fundamental), with right (left) transformations,
$\theta_g \psi^{\dagger}_m \theta^{\dagger}_g = \psi^{\dagger}_{m'} D_{m'm}\left(g\right)$
($\widetilde\theta_g \psi^{\dagger}_m \widetilde\theta^{\dagger}_g = D_{mm'}\left(g\right)\psi^{\dagger}_{m'} $);
$D_{mn}^{j}\left(g\right)$ is the $j$ irreducible  representation of $g$,  $j$ is omitted when the representation of $\psi^{\dagger}_{m}$ is used; throughout this paper, summation of doubled indices is assumed.
 We will consider here a staggered fermionic picture \cite{Susskind1977}, splitting the vertices into two sublattices -
even ($e$) and odd ($o$), after a particle-hole transformation of $o$ \cite{Zohar2015b,Zohar2016a}.
Global transformations will usually involve two different types of transformations (right/left) for the two sublattices. For that we define $\check{\theta}\left(\mathbf{x}\right)$,
which is $\theta$ for $\mathbf{x}\in e$ and $\widetilde\theta^{\dagger}$ for $\mathbf{x}\in o$.

 The gauge symmetry is invariance under gauge transformations:
\begin{equation}
\hat\Theta_g\left(\mathbf{x}\right) = \underset{k=1...d}{\prod}\left(\widetilde{\Theta}_g\left(\mathbf{x},k\right)\Theta^{\dagger}_g\left(\mathbf{x-e}_k,k\right)\right)
\check{\theta}^{\dagger}_g\left(\mathbf{x}\right),
\label{gaugetrans}
\end{equation}
involving a vertex and all the links starting and ending there ($\mathbf{e}_k$ is a unit vector in the $k$ direction).
A gauge-invariant state $\left|\Psi\right\rangle$ satisfies $\hat\Theta_g\left(\mathbf{x}\right)\left|\Psi\right\rangle=\left|\Psi\right\rangle$ for any $\mathbf{x}\in\mathbb{Z}^d,g\in G$ (disregarding static charges,
which we do not discuss here).

\emph{Constructing  gaussian fermionic states with global symmetry}. Fermionic gaussian states are completely characterized by their covariance
matrix \cite{Bravyi05} - a matrix of correlations, composed out of the fundamental blocks $\mathcal{Q}_{mn}\left(\mathbf{x},\mathbf{y}\right) = i\left\langle\left[\psi_m\left(\mathbf{x}\right),\psi_n\left(\mathbf{y}\right)\right]\right\rangle/2$ and
$\mathcal{R}_{mn}\left(\mathbf{x},\mathbf{y}\right) = i\left\langle\left[\psi_m\left(\mathbf{x}\right),\psi^{\dagger}_n\left(\mathbf{y}\right)\right]\right\rangle/2$. For such states, in particular for fermionic gaussian PEPS,
the determination of norms and expectation values can be done very efficiently, using the covariance matrix formalism \cite{Bravyi05,Kraus2010} (see App. \ref{appd}).

 Following \cite{Zohar2015b,Zohar2016a},
 we start by constructing the state $\left|\psi_0\right\rangle$, which is globally invariant under $G$ transformations. On top of
 the physical fermionic modes $\psi^{\dagger}_m\left(\mathbf{x}\right)$ at each vertex, we  introduce
\emph{virtual fermionic modes} - $\left\{  c^{j,\alpha\dagger}_m\left(\mathbf{x},\pm k\right)   \right\}$, corresponding to the outgoing (+) or ingoing (-) links in direction $k$ (Fig. \ref{fig1}).
These are spinors that undergo transformations with respect to the $j$ representation of the group, with operators $\theta_g,\widetilde\theta_g$ as the ones defined above for the physical fermions. $m$ labels the spinor's components. All the $\text{dim}\left(j\right)$  components of any  $j$ used must be included; several copies of the same representation may also be used, which is labeled by $\alpha$.
This generalizes the minimalistic constructions of \cite{Zohar2015b,Zohar2016a}, that involved spinors belonging to the same representations as the physical fermions \cite{Zohar2015b,Zohar2016a}.

Out of the fermionic transformation operators, both physical and virtual,
we contruct a \emph{virtual gauge transformation},
$\hat\theta_g=\underset{k=1...d}{\prod}\left(\widetilde\theta_g^{+k}\theta_g^{\dagger-k}\right)\theta_g^{\psi\dagger}$
- similar to the physical gauge transformation (\ref{gaugetrans}), but involving \emph{virtual gauge fields} rather than physical ones. This will be useful when we gauge the state and introduce the physical gauge degrees of freedom.

At each vertex $\mathbf{x}$ we introduce a gaussian operator $A\left(\mathbf{x}\right)=\text{exp}\left(T_{ij}a^{\dagger}_i\left(\mathbf{x}\right) a^{\dagger}_j\left(\mathbf{x}\right)\right)$,
where $a^{\dagger}_i$ are fermionic creation operators (physical or virtual). The matrix $T_{ij}$ is determined by demanding that the operator is a $G$ singlet:
\begin{equation}
\hat\theta_g\left(\mathbf{x}\right) A\left(\mathbf{x}\right) \hat\theta_g^{\dagger}\left(\mathbf{x}\right) = A\left(\mathbf{x}\right)
\label{transver}
\end{equation}
Note that $A$, as a singlet, will satisfy a similar equation if left and right transformations are exchanged.

Next, we introduce the operators
$V\left(\mathbf{x},k\right) = \text{exp}\left(B_{kl}c^{\dagger}_k\left(\mathbf{x},k\right)c^{\dagger}_l\left(\mathbf{x+e}_k,-k\right)\right)$ connecting virtual modes on both edges of a link;
the projector for the empty state on a link, $\Omega\left(\mathbf{x},k\right)$; and the unnormalized projectors
$\omega\left(\mathbf{x},k\right) = V\left(\mathbf{x},k\right) \Omega\left(\mathbf{x},k\right) V^{\dagger}\left(\mathbf{x},k\right)$ (Fig. \ref{fig1}).
The coefficients $B_{mn}$ are determined by demanding the symmetry condition
\begin{equation}
\theta_g^{k+}\left(\mathbf{x}\right)\theta_g^{k-}\left(\mathbf{x+e}_k\right)\omega\left(\mathbf{x},k\right)\theta_g^{k+\dagger}\left(\mathbf{x}\right)\theta_g^{k-\dagger}\left(\mathbf{x+e}_k\right)=\omega\left(\mathbf{x},k\right)
\label{translink}
\end{equation}
which will automatically hold for left transformations as well.

As in \cite{Zohar2015b,Zohar2016a}, $\left|\psi_0\right\rangle$ is obtained by acting with a proper product of  operators on the Fock vacuum (physical and virtual) $\left|\Omega\right\rangle$:
$\left|\psi_0\right\rangle = \underset{\mathbf{x},k}{\prod}\omega\left(\mathbf{x},k\right) \underset{\mathbf{x}}{\prod} A\left(\mathbf{x}\right) \left|\Omega\right\rangle$.

Gaussian operators are exponentials of quadratic functions of mode operators, and they keep this property when they are multiplied.
Gaussian states are those that can be written as a gaussian operator acting on the vacuum. Since $\Omega\left(\mathbf{x},k\right)$ and $V\left(\mathbf{x},k\right)$ are gaussian, so are
$\omega\left(\mathbf{x},k\right)$. Furthermore, since $A$ is gaussian as well, the state $\left|\psi_0\right\rangle$ is gaussian.
It is a product of a many-body state of the physical fermions, with the states created by the operators $V\left(\mathbf{x},k\right)$ on the links.
$\left|\psi_0\right\rangle$ is invariant under global transformations of the form
$\underset{\mathbf{x}}{\prod}\check{\theta}^{\dagger}_{h}\left(\mathbf{x}\right)$ on the physical  fermions.
This can be seen using the symmetry properties (\ref{transver},\ref{translink}).

\emph{Gauging.} We proceed with gauging the globally invariant states, generalizing \cite{Zohar2016a}, and making the symmetry local by the introduction of gauge fields on the links.
 On each link,  we define the \emph{group element operators} \cite{KogutSusskind},
$U^{j}_{mn}\left(\mathbf{x},k\right) = \int dg\left(\mathbf{x},k\right) \left|g\left(\mathbf{x},k\right)\right\rangle \left\langle g\left(\mathbf{x},k\right)\right| D^{j}_{mn}\left(g\right)$ (see App. \ref{appa}), transforming as the $j$ representation under group transformations.

Gauging is done by making the replacements
 $c_m^{j,\alpha\dagger}\left(\mathbf{x},+k\right) \rightarrow U^{j}_{mn}\left(\mathbf{x},k\right)c_n^{j,\alpha\dagger}\left(\mathbf{x},+k\right)$ for $\mathbf{x} \in e$, ($\overline{U}^{j}_{mn}\left(\mathbf{x},k\right)c_n^{j,\alpha\dagger}\left(\mathbf{x},+k\right)$ for $\mathbf{x} \in o$) in $A$, and acting on a product of gauge field trivial states on the links. The operator $U$ plays the role of a rotation matrix (whose elements are operators acting on the gauge field) applied to the virtual spinor, mixing its components, but not within different copies ($\alpha$); moreover, $U^{j}$ is the same for every $\alpha$ copy of the virtual fermions.

Our goal now is to show that we can write the gauged state in the form (\ref{psibig}), where $\left|\Psi\left(\mathcal{G}\right)\right\rangle$ is a gaussian state. As ${U}^{j}$ is a unitary matrix, gauging is the result of a unitary transformation: we define the gaussian operator  $\mathcal{U}_g\left(\mathbf{x},k\right)=\widetilde\theta^{k+}_g\left(\mathbf{x}\right)$  for $\mathbf{x}\in e$ ($\theta^{k+\dagger}_g\left(\mathbf{x}\right)$ for $\mathbf{x} \in o$ ), and construct a unitary operator that entangles the gauge field and virtual
fermions $c^{j,\alpha}_n\left(\mathbf{x},+k\right)$ on a link's beginning:
$\mathcal{U}\left(\mathbf{x},k\right) = \int dg\left(\mathbf{x},k\right)\ \left|g\left(\mathbf{x},k\right)\right\rangle\left\langle g\left(\mathbf{x},k\right)\right|\otimes \mathcal{U}_g\left(\mathbf{x},k\right)$
 (see Fig. \ref{fig1}). With this we can finally write
 \begin{equation}
\left|\Psi\right\rangle = \left|G\right|^{N_{\text{links}}/2} \underset{\mathbf{x},k}{\prod}\omega\left(\mathbf{x},k\right) \underset{\mathbf{x},k}{\prod}\mathcal{U}\left(\mathbf{x},k\right)\left|000\right\rangle_{\mathbf{x},k}
\underset{\mathbf{x}}{\prod} A\left(\mathbf{x}\right) \left|\Omega\right\rangle
\label{gaugedstate}
\end{equation}
  which is gauge-invariant ($\hat\Theta_g\left(\mathbf{x}\right)\left|\Psi\right\rangle = \left|\Psi\right\rangle \quad \forall g\in G,\mathbf{x}\in\mathbb{Z}^d$; the fermionic construction truncates the physical link Hilbert spaces -  see App. \ref{appb} for details).

Finally, we express $\left|\Psi\right\rangle$ in terms of group element states. Using $\left\langle g |000\right\rangle = \left|G\right|^{-1/2}$ (see app. \ref{appa}), we obtain the form (\ref{psibig}),
 with
\begin{equation}
\left|\psi\left(\mathcal{G}\right)\right\rangle = \underset{\mathbf{x},k}{\prod}\omega\left(\mathbf{x},k\right) \underset{\mathbf{x},k}{\prod}\mathcal{U}_{g\left(\mathbf{x},k\right)}\left(\mathbf{x},k\right) \underset{\mathbf{x}}{\prod} A\left(\mathbf{x}\right) \left|\Omega\right\rangle
\label{psiG}
\end{equation}
As for any $g\left(\mathbf{x},k\right)$  $\mathcal{U}_{g\left(\mathbf{x},k\right)}\left(\mathbf{x},k\right)$ is gaussian, we conclude that
$\left|\psi\left(\mathcal{G}\right)\right\rangle$ is a gaussian state
(of fermions coupled to a static field $\mathcal{G}$ - see App. \ref{appc}).

\emph{Efficient computation of expectation values.} Finally, we would like to see how (\ref{psibig}) allows us to perform efficient calculations for these states, i.e. to combine Monte-Carlo methods with fermionic GGPEPS, without facing the sign problem. As
$U^{j}_{mn}\left|g\right\rangle=D^{j}_{mn}\left(g\right)\left|g\right\rangle$, $\left|\mathcal{G}\right\rangle$ are eigenstates of any function of the group element operators:
\begin{equation}
F\left(\left\{U^{j}_{mn}\left(\mathbf{x},k\right)\right\}\right)\left|\mathcal{G}\right\rangle = F\left(\left\{D^{j}_{mn}\left(g\left(\mathbf{x},k\right)\right)\right\}\right)\left|\mathcal{G}\right\rangle
\label{eigst}
\end{equation}
which will allow us to express the expectation values of observables in a very convenient way.

First, let us consider \emph{Wilson Loops},
$W\left(C\right)=\text{Tr}\left(\underset{\left\{\mathbf{x},k\right\}\in C}{\prod}U\left(\mathbf{x},k\right)\right)$,
 where $C$ is some closed oriented path (the matrices $U$ should be
replaced by $U^{\dagger}$ according to the orientation of the respective link along $C$). These gauge-invariant operators are important, as, for example, they serve as  order parameters for static charges
confinement in pure-gauge theories  \cite{Wilson,Fradkin1978}.

Using the inner product of configuration states $\left|\mathcal{G}\right\rangle$ and (\ref{eigst}), one simply obtains
$\left\langle W\left(C\right) \right\rangle
=\int \mathcal{DG} F_{W\left(C\right)}\left(\mathcal{G}\right) p\left(\mathcal{G}\right)$
where
$F_{W\left(C\right)}\left(\mathcal{G}\right)=\text{Tr}\left(\underset{\left\{\mathbf{x},k\right\}\in C}{\prod}D\left(g\left(\mathbf{x},k\right)\right)\right)$ and
\begin{equation}
p\left(\mathcal{G}\right) = \frac{\left\langle\psi\left(\mathcal{G}\right)|\psi\left(\mathcal{G}\right)\right\rangle}{ \int \mathcal{D}\mathcal{G}' \left\langle\psi\left(\mathcal{G}'\right)|\psi\left(\mathcal{G}'\right)\right\rangle}
\end{equation}
As $\left\langle\psi\left(\mathcal{G}\right)|\psi\left(\mathcal{G}\right)\right\rangle$ is the square of a norm, $p\left(\mathcal{G}\right)$ is a probability density, and hence $\left\langle W\left(C\right) \right\rangle$ may computed by Monte-Carlo methods. Furthermore, $p\left(\mathcal{G}\right)$ can be efficiently computed, since $\left|\psi\left(\mathcal{G}\right)\right\rangle$ is gaussian \cite{Bravyi05} (see App. \ref{appd}).

The expectation value of other types of observables, e.g. meson operators involving both gauge field and fermionic operators, as well as local operators not diagonal in group element terms, could also be comoputed with similar methods (see  App. \ref{appe} for details).

\emph{Illustration.} To demonstrate our method, we performed some Wilson loop computations for a $2+1$d $\mathbb{Z}_3$ pure gauge theory (with translational and rotational invariance), with a low numerical component, sufficient for illustration purposes. This was done using the $U(1)$ parameterization of \cite{Zohar2015b}. The fermionic construction imposes a three-dimensional truncation of the physical (link) Hilbert spaces (see app. \ref{appf} for more
details). As $\mathbb{Z}_3$ is a subgroup of $U(1)$, the
PEPS is also invariant under it, and thus could be used as a non-truncated $\mathbb{Z}_3$ state.
In this case,
the Monte-Carlo integration on a link reduces to summing over the three group elements $\left\{e^{2\pi i q/3}\right\}_{q=-1}^{1}$.

\begin{figure}[t!]
  \centering
  \includegraphics[width=0.5\textwidth]{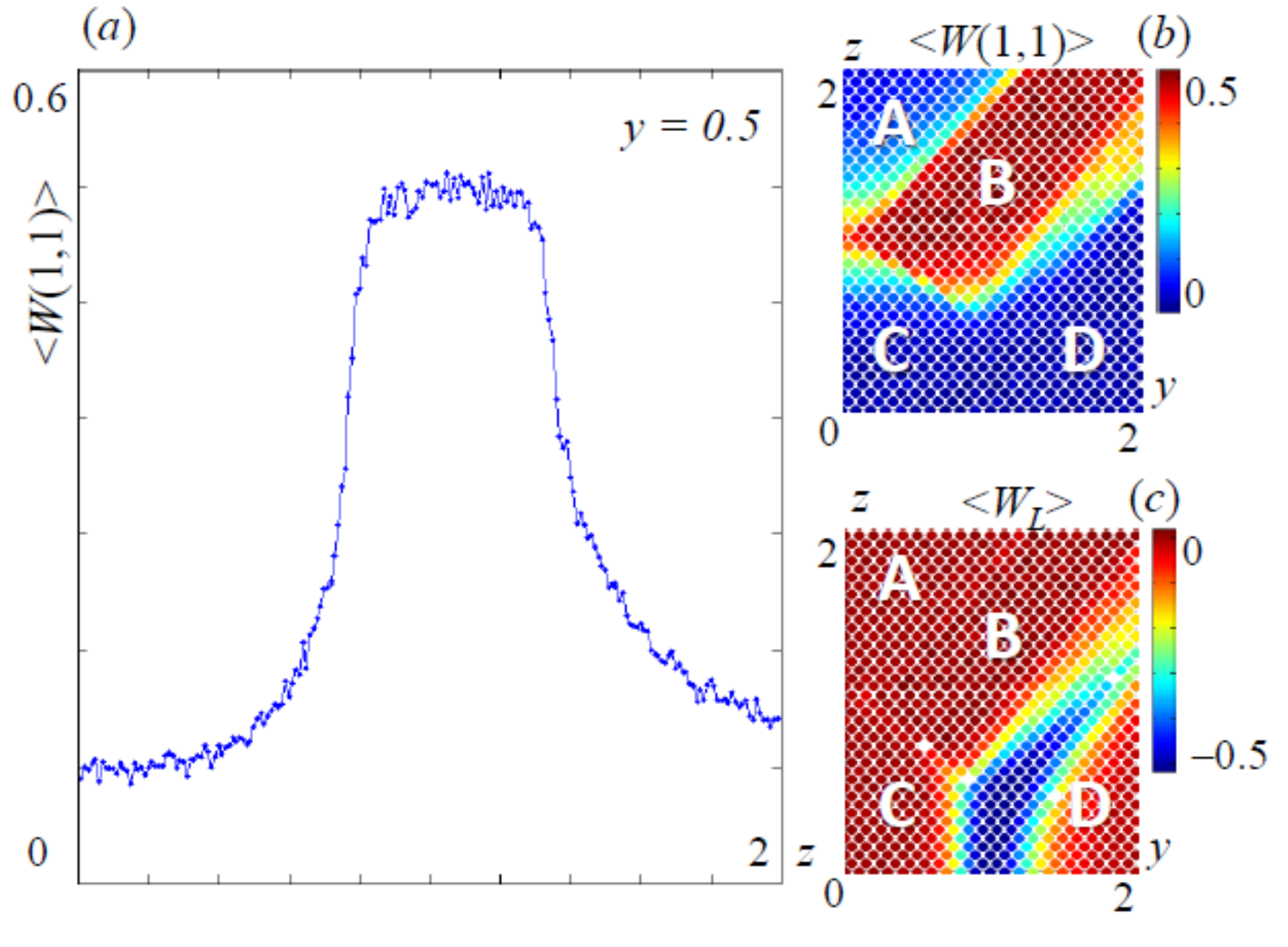}\\
  \caption{Wilson loop results for $\mathbb{Z}_3$ in $2+1$d for different values of the parameters $y,z$ upon which the state depends. (a) a single plaquette, along $y=0.5$. One can see the transitions between the phases C,B and A. (b) a single plaquette, for $y,z\in\left[0,2\right]$. The
phase boundaries are clearly seen, besides the one between C and D, as this observable is insensitive to it: this is a region in parameter space dominated by $y$, responsible for straight
flux lines rather than corners (see App. \ref{appf}), of which the single plaquette is built. (c) A similar plot for the Wilson loop (or line) winding around the horizontal circumference. The C,D transition is seen,
as a line depends on $y$ and not on $z$. The boundaries on the left, $z$ dominated region, are not seen for a similar reason. The phase labels are taken from \cite{Zohar2015b}.}\label{fig2}
\end{figure}

The state involves only virtual fermions - no physical ones (as it is pure-gauge). Therefore the $A$ operators are only used for connecting the gauge field Hilbert spaces on the links in a gauge-invariant way. It depends on two parameters, $y$ and $z$. For pure gauge theories, the calculation of $\left\langle \psi \left(\mathcal{G}\right) | \psi \left(\mathcal{G}\right) \right\rangle$ is significantly simplified (see App. \ref{appd}) for the explicit form). We computed, as functions of these two parameters, the expectation value of $\mathbb{Z}_3$ loop operators -
of a single plaquette $\left\langle W(1,1)\right\rangle = \left\langle U\left(0,1\right)U\left(\mathbf{e}_1,2\right)U^{\dagger}\left(\mathbf{e}_2,1\right)U^{\dagger}\left(0,2\right)\right\rangle$, and the horizontal circumference
$\left\langle W_L\right\rangle = \left\langle\overset{7}{\underset{m=0}{\prod}}U\left(m\mathbf{e}_1,1\right)\right\rangle$  - for $y,z \in \left[0,2\right]$, in a $8\times8$
periodic system. The results show apparent jumps in the expectation value at the expected phase boundaries, which were found using other methods in \cite{Zohar2015b}. Note that here we computed the expectation value for a $\mathbb{Z}_3$ theory and in \cite{Zohar2015b}
it was $U(1)$,
and therefore the physics of the phases may be different; however, the phase boundaries, for a PEPS, are properties of the state, and as the state here is the same one used in \cite{Zohar2015b}, the same phase boundaries should be, and have
been,
found, as can be seen in Fig. \ref{fig2}.

\emph{Conclusion and Outlook.} In this paper, we first generalized the fermionic GGPEPS constructed in \cite{Zohar2015b,Zohar2016a} to higher bond dimensions. Those are free fermion (gaussian) states for the matter, which are gauged with the standard procedure \cite{Zohar2015b,Zohar2016a}. Then, we showed that these fermionic GGPEPS could be
reexpressed, using group element states, as an expansion in products of gauge field configurations and fermionic gaussian states that depend on the gauge fields as parameters (background fields). One could use that for writing the expectation values of gauge-invariant operators for these states in a form that allows an efficient Monte-Carlo calculation, not in a Euclidean spacetime, in which the fermionic part could be efficiently contracted using the gaussian formalism, circumventing the sign problem. We have demonstrated the method with a simple, proof of principle illustration of probing the phase boundaries for a $\mathbb{Z}_3$ PEPS in $2+1$d. The method can then be applied for the study of HEP problems, suggesting GGPEPS as a tool for numerical, variational studies of LGTs in larger systems and higher dimensions. Furthermore, the method can be easily combined with other techniques (string-bond, entangled plaquette, or neural networks) by just adding a function $f\left(\mathcal{G}\right)$ inside the integral (\ref{psibig}) such that it can be easily computed \cite{Glasser2017}.

\begin{acknowledgments}
 EZ would like to thank  I. Glasser and N. Hallakoun for the technical support. JIC is partially supported by the EU, ERC grant QUENOCOBA 742102.
 \end{acknowledgments}

\appendix
\section{More details about the Hilbert space and the representation basis}\label{appa}

In this work, we used the \emph{group element states} $\left\{\left|g\right\rangle\right\}_{g\in G}$ for the local (link) Hilbert spaces of the gauge field, and defined
 the group transformations, right $\Theta_g\left|h\right\rangle = \left|hg^{-1}\right\rangle$ and left $\widetilde{\Theta}_g\left|h\right\rangle = \left|g^{-1}h\right\rangle$ on them.

One may also use
the \emph{representation basis} \cite{Zohar2015},  whose states $\left|jmn\right\rangle$ are labeled by an irreducible representation $j$ and two identifiers within the representation, $m$ and $n$,
corresponding to left and right degrees of freedom.
The transition from $\left|g\right\rangle$ to $\left|jmn\right\rangle$ is given by
\begin{equation}
\left\langle g | jmn \right\rangle = \sqrt{\frac{\text{dim}\left(j\right)}{\left|G\right|}} D^{j}_{mn}\left(g\right)
\label{changebasis}
\end{equation}
which is simply a generalization of Wigner's formula for the eigenfunctions of the isotropic rigid rotator (Wigner matrices) \cite{Rose1995,Edmonds1996}.

To understand better the $m$ and $n$ notions, let us use (\ref{changebasis}) to see how these states transform under the group:
\begin{equation}
\Theta_g \left|jmn\right\rangle = \left|jmn'\right\rangle D^{j}_{n'n}\left(g\right)\quad  \quad\tilde{\Theta}_g\left|h\right\rangle = D^{j}_{mm'}\left(g\right)\left|jm'n\right\rangle
\end{equation}
One particular state in this basis is the singlet state - $\left|000\right\rangle$, corresponding to the trivial representation. It is invariant under any group transformation.
This is the only representation state we use in the main text, as it is used for the gauging procedure. Note that (\ref{changebasis}) implies that
\begin{equation}
\left\langle g |000\right\rangle = \left|G\right|^{-1/2}
\end{equation}
which was used by us in the main text.

We also introduced the \emph{group element operators} \cite{Zohar2015},
\begin{equation}
U^{j}_{mn} = \int dg D_{mn}^{j}\left(g\right)\left|g\right\rangle\left\langle g \right|.
\label{Udef}
\end{equation}
These are matrices of operators: the matrix indices, $m,n$, refer to a linear space called either group, color or gauge space, on which the group transformations act. Each such matrix element is an operator on the local
 Hilbert space on the link.
It is clear from the definition that the different matrix elements of $U^{j}_{mn}$ commute, and hence one may define functions of these operators as if they were matrices of numbers.
Note that
\begin{equation}
\left|jmn\right\rangle=\sqrt{\text{dim}\left(j\right)}U^{j}_{mn}\left|000\right\rangle.
\end{equation}

In the main text, we stated that local gauge symmetry is simply invariance under the gauge transformations
\begin{equation}
\hat\Theta_g\left(\mathbf{x}\right) = \underset{k=1...d}{\prod}\left(\widetilde{\Theta}_g\left(\mathbf{x},k\right)\Theta^{\dagger}_g\left(\mathbf{x}-\hat{\mathbf{k}},k\right)\right)
\check{\theta}^{\dagger}_g\left(\mathbf{x}\right)
\end{equation}
involving a vertex and all the links starting and ending there.
A gauge invariant state $\left|\Psi\right\rangle$ satisfies $\hat\Theta_g\left(\mathbf{x}\right)\left|\Psi\right\rangle=\left|\Psi\right\rangle$ for each $\mathbf{x}\in\mathbb{Z}^d,g\in G$.

If $G$ is a  Lie group, one may define its left and right generators, $L_a,R_a$ respectively, satisfying the group's algebra
\begin{equation}
\begin{aligned}
&\left[R_a,R_b\right]=if_{abc}R_c \\
&\left[L_a,L_b\right]=-if_{abc}L_c \\
&\left[R_a,L_b\right]=0
\end{aligned}
\end{equation}
as well as the matrix $j$ represenation of the generators, $T^j_a$, with
\begin{equation}
\left[T^j_a,T^j_b\right]=if_{abc}T^j_c \\
\end{equation}
where $f_{abc}$ are the group's structure constants.

These can be used for expressing the transformation operators, as well as the representation matrices, using the group parameters $\phi_a\left(g\right)$:
\begin{equation}
\begin{aligned}
\Theta_g &= e^{i\phi_a\left(g\right) R_a} \\
\widetilde\Theta_g &= e^{i\phi_a\left(g\right)  L_a}\\
D^{j}\left(g\right) &= e^{i \phi_a\left(g\right) T_a}
\end{aligned}
\end{equation}
Formally, one may also define operators $\hat\phi_a$,
such that
\begin{equation}
U^j_{mn} = \left(e^{i \hat\phi_a T_a}\right)_{mn}
\end{equation}
$\hat\phi_a$ play the role of the vector potential on a link, which is not a well defined quantity on a lattice (where one uses the group elements instead of its algebra). Therefore
the group element operator is the lattice analog to a Wilson line along a link.

As transformation generators, the $R,L$ operators satisfy
\begin{equation}
\begin{aligned}
\left[R_a,U^{j}_{mn}\right]&=U^{j}_{mn'}\left(T_a\right)^{j}_{n'n} \\
\left[L_a,U^{j}_{mn}\right]&=\left(T_a\right)^{j}_{mm'} U^{j}_{m'n}
\end{aligned}
\end{equation}

It is also possible to express the gauge transformation in this way, and define
\begin{equation}
\hat\Theta_g\left(\mathbf{x}\right) = e^{i \phi_a G_a}
\end{equation}
with
\begin{equation}
G_a\left(\mathbf{x}\right) = \underset{k=1...d}{\sum}\left(L_a\left(\mathbf{x},k\right) - R_a\left(\mathbf{x-\hat{k}},k\right)\right)-Q_a\left(\mathbf{x}\right)
\end{equation}
where $Q_a\left(\mathbf{x}\right)$ are the fermionic charges (see, e.g. \cite{Zohar2015,Zohar2015b,Zohar2016a}). Then, gauge invariance (without static charges) implies
\begin{equation}
G_a\left(\mathbf{x}\right)\left|\Psi\right\rangle = 0 \quad \forall \mathbf{x},g,a
\end{equation}
 This equation is
known as the Gauss law, and interprets $R_a,L_a$ as the (right and left) electric fields.

\section{Gauging in the representation basis and the truncation of the physical Hilbert spaces}\label{appb}

We mentioned in the main text that the fermionic construction imposes a truncation of the gauge field physical Hilbert space. Full details may be found in \cite{Zohar2015b,Zohar2016a}; here we shall explain that briefly.
In the gauging procedure, we defined the following unitary operators, that entangle the gauge field and virtual fermions on a link:
\begin{equation}
\mathcal{U}\left(\mathbf{x},k\right) = \int dg \left|g\left(\mathbf{x},k\right)\right\rangle\left\langle g\left(\mathbf{x},k\right)\right|\otimes \mathcal{U}_g\left(\mathbf{x},k\right)
\end{equation}
 (see Fig. \ref{fig1}), leading us to the gauged state
\begin{equation}
\left|\Psi\right\rangle =\left|G\right|^{N_{\text{links}}/2} \underset{\mathbf{x},k}{\prod}\omega\left(\mathbf{x},k\right) \underset{\mathbf{x},k}{\prod}\mathcal{U}\left(\mathbf{x},k\right)\left|000\right\rangle_{\mathbf{x},k}
\underset{\mathbf{x}}{\prod} A\left(\mathbf{x}\right) \left|\Omega\right\rangle
\label{GS}
\end{equation}

The transformation $\mathcal{U}\left(\mathbf{x},k\right)$, when acting on the virtual creation operators of $A\left(\mathbf{x}\right)$, simply rotates them, within $A$, by the matrix $U$ ($\overline{U}$) on the link, for an even (odd)
 $\mathbf{x}$:
 $c_m^{j,\alpha\dagger}\left(\mathbf{x},+k\right) \rightarrow U^{j}_{mn}\left(\mathbf{x},k\right)c_n^{j,\alpha\dagger}\left(\mathbf{x}+k\right)$ for $\mathbf{x} \in e$, (or $\overline{U}^{j}_{mn}\left(\mathbf{x},k\right)c_n^{j,\alpha\dagger}\left(\mathbf{x},+k\right)$ for $\mathbf{x} \in o$).
In other words, the physical electric field is identified with a virtual electric field, defined by the virtual fermions. The action $U^{j}_{mn}\left(\mathbf{x},k\right)c_n^{j,\alpha\dagger}\left(\mathbf{x}+k\right)$ (and multiple actions thereof, as such operators
appear in the exponential of $A$) on the product of fermionic vacuum and gauge field singlet, excite both the virtual and physical electric fields in a correlated way. However, the virtual fields are truncated, as they are created
from a finite set of fermionic operators, which truncates the physical Hilbert space on the link as well. The truncation is done in representation basis: the physical gauge field states on a link are created from the singlet
$\left|000\right\rangle$ with products of $U$ ($\overline{U}$) matrix elements, accompanied by virtual fermionic operators, which due to the fermionic statistics, impose the truncation.

For example, in the $U(1)$ case of \cite{Zohar2015b},
 there are two virtual fermionic modes on each edge, corresponding to the representations $j=\pm1$, that may lead together to the total representations $0,\pm1$ - virtual electric field configurations, with $0$ corresponding
to no fermions or both present (the fermionic statistics forces the creation of a singlet), and $\pm1$ to the presence of a single fermion. This truncates the physical Hilbert space on the link, making it three dimensional,
with electric field $0,\pm1$ (not differentiating between the two possible ways to obtain a virtual zero field). As $\mathbb{Z}_3$ is a subgroup of $U(1)$, the
PEPS is also invariant under it, and in general, one could use the same state $\left|\Psi\right\rangle$ for studying $U(1)$ models with an electric field truncation $\left|E\right|\leq\ell$ as well as $\mathbb{Z}_{2\ell+1}$.
The difference between the two cases will arise for the observables whose expectation values and correlations are computed - i.e., whether they respect the $U(1)$ symmetry or only that of the subgroup $\mathbb{Z}_{2\ell+1}$.

In the $SU(2)$ case of \cite{Zohar2016a}, once again there are two virtual modes per edge,
corresponding to the two spin half states. These may realize only the representations $0,1/2$ on the link (1 is prevented by the fermionic statistics), and the Hilbert space of the link is truncated to
 $\left|jmn\right\rangle$ states with $j=0,1/2$ - a five dimensional space. In this case, however, we could not use the same state construction for studying a subgroup, as there is no  non-Abelian 5 dimensional subgroup of $SU(2)$.

\section{More on the transformation properties of gauged fermionic gaussian states}\label{appc}

In the main text, we wrote that the fermionic gaussian state $\left|\psi\left(\mathcal{G}\right)\right\rangle$ describes fermions coupled to a static background field $\mathcal{G}$. Let us see why.
Using Eqs. (\ref{transver},\ref{translink}), we obtain that
\begin{equation}
\underset{\mathbf{x}}{\prod}\check{\theta}^{\dagger}_{h\left(\mathbf{x}\right)}\left(\mathbf{x}\right)
 \left|\psi\left(\mathcal{G}\right)\right\rangle = \left|\psi\left(\mathcal{G'}\right)\right\rangle
 \end{equation}
 where $\mathcal{G'} = \left\{g'\left(\mathbf{x},k\right) = h^{-1}\left(\mathbf{x}\right)g\left(\mathbf{x},k\right)h\left(\mathbf{x+\hat{k}}\right)\right\}$: under fermionic transformations with arbitrary, position-dependent group elements $\left\{h\left(\mathbf{x}\right)\right\}$,
the gauge field configuration transforms as $\mathcal{G} \rightarrow \mathcal{G'} = \left\{g'\left(\mathbf{x},k\right) = h^{-1}\left(\mathbf{x}\right)g\left(\mathbf{x},k\right)h\left(\mathbf{x+\hat{k}}\right)\right\}$:
 $\left|\psi\left(\mathcal{G}\right)\right\rangle$ transforms, indeed, as a state with a background field configuration $\mathcal{G}$. Note, that as the group elements used as parameters for this transformation
are vertex-dependent, the transformation is local, and could be performed similarly only on few vertices (or one), in all cases giving rise to a physically equivalent state.

This also implies the non-surprising result, that the globally-invariant state corresponds to a state without background field, or, in other words, in which $\mathcal{G}$ is the identity element ($e$) everywhere:
\begin{equation}
\left|\psi_0\right\rangle = \left|\psi\left(e\right)\right\rangle
\end{equation}

The state $\left|\Psi\right\rangle$ is gauge invariant by construction, as shown in \cite{Zohar2015b,Zohar2016a}. Here, however, we shall give an alternative proof for that, using the group element states, and the physical interpretation of  $\left|\psi\left(\mathcal{G}\right)\right\rangle$ presented above. Let us apply a local gauge transformation with a group element $h$ at the vertex $\mathbf{x}$ and use the transformation properties of $\left|\psi\left(\mathcal{G}\right)\right\rangle$ and $\left|\mathcal{G}\right\rangle$ :
\begin{widetext}
\begin{equation}
\hat\Theta_h\left(\mathbf{x}\right)\left|\Psi\right\rangle =
\int \mathcal{DG} \underset{k=1...d}{\prod}\left(\widetilde{\Theta}_h\left(\mathbf{x},k\right)\Theta^{\dagger}_h\left(\mathbf{x-e}_k,k\right)\right)\left|\mathcal{G}\right\rangle
\check{\theta}^{\dagger}_h\left(\mathbf{x}\right)\left|\psi\left(\mathcal{G}\right)\right\rangle
=\int \mathcal{DG} \left|\mathcal{G'}\right\rangle \left|\psi\left(\mathcal{G'}\right)\right\rangle  =\int \mathcal{DG'} \left|\mathcal{G'}\right\rangle \left|\psi\left(\mathcal{G'}\right)\right\rangle= \left|\Psi\right\rangle
\end{equation}
\end{widetext}
where we have used the invariance of the integration measure under a unitary coordinate change  - as the gauge transformation is.

\section{The gaussian formalism}\label{appd}

Fermionic gaussian states are fully characterized by their covariance matrix \cite{Bravyi05}. In the main text, we defined it for the physical fermions, but obviously it could be extended for the virtual modes as well. Besides that, the gaussian formalism becomes extremely simple when one, instead of using Dirac fermions, uses a majorana formulation - i.e., for every fermionic mode $a_i$, define the two hermitian Majorana operators
\begin{equation}
\gamma_i^{(1)}=\left(a_i+a_i^{\dagger}\right) \quad ; \quad \gamma_i^{(2)}=i\left(a_i-a_i^{\dagger}\right)
\end{equation}
If we unite all the $2N$ Majorana modes of a system containing $N$ fermionic modes under $\left\{\gamma_a\right\}_{a=1}^{2N}$, we can write that they satisfy the algebra
\begin{equation}
\left\{\gamma_a,\gamma_b\right\} = 2 \delta_{ab}
\end{equation}
and define the covariance matrix for a gaussian state $\left|\phi\right\rangle$ in Majorana terms
\begin{equation}
\Gamma_{ab} = \frac{i}{2}\left\langle  \left[\gamma_a,\gamma_b\right] \right\rangle
=\frac{i}{2} \frac{\left\langle \phi \right| \left[\gamma_a,\gamma_b\right] \left|\phi\right\rangle}{\left\langle \phi |\phi\right\rangle}
\end{equation}

To obtain the covariance matrix of the state $\left|\psi\left(\mathcal{G}\right)\right\rangle$, one can use a gaussian map \cite{Bravyi05,Kraus2010}. This is done as follows. Define the state
\begin{equation}
\left|A\right\rangle = \underset{\mathbf{x}}{\prod} A\left(\mathbf{x}\right) \left|\Omega\right\rangle
\end{equation}
and denote its density matrix by $\rho_A$. It is a gaussian product state, that does not introduce mixing among different vertices. Thus, its covariance matrix $M$ will be a direct sum of the covariance matrices of each vertex, $M\left(\mathbf{x}\right)$
\begin{equation}
M = \underset{\mathbf{x}}{\bigoplus}M\left(\mathbf{x}\right)
\end{equation}
and in the translationally invariant case, one will simply have that $M\left(\mathbf{x}\right) = M_0$.

We can thus express $\left|\psi_0\right\rangle$ as
\begin{equation}
\left|\psi_0\right\rangle = \underset{\mathbf{x},k}{\prod}\omega\left(\mathbf{x},k\right) \left|A\right\rangle
\end{equation}
If we denote the density matrix corresponding to the unnormalized operators $\omega\left(\mathbf{x},k\right)$ by $\rho_B$, we can write the density matrix of physical fermions corresponding to $\left|\psi_0\right\rangle$ as
\begin{equation}
\rho_{0} = \text{Tr}_{V}\left( \rho_B \rho_A\right)
\end{equation}
which involves a fermionic partial trace on the virtual modes, that has to be carefully defined \cite{Bravyi05}.

We reorder the covariance matrix $M$ such that it has the following form:
\begin{equation}
M = \left(
  \begin{array}{cc}
    M_A & M_B \\
    -M^T_B & M_D \\
  \end{array}
\right)
\end{equation}
where $M_A$ is a block that corresponds to correlations of physical fermions with themselves, $M_D$ corresponds to the same for virtual fermions, and $M_B$ is for mixed correlations. We also construct the covariance matrix
$\Gamma_{\text{in}}$ corresponding to $\rho_B$; its dimension is equal to this of $M_D$, as it only involves virtual fermions, and we order the matrix in the same order of the virtual modes in $M$. Then, the covariance matrix
of the output, physical state $\left|\Psi\right\rangle$ is given by \cite{Bravyi05}
\begin{equation}
\Gamma_{\text{out}} = M_A + M_B\left(M_D - \Gamma_{\text{in}}\right)^{-1}M_B^T
\end{equation}
(this holds only if $\rho_B$ is pure, but this is our case here). If the PEPS is translationally invariant, one can decompose everything into momentum blocks using a Fourier transform \cite{Kraus2010}, but we are interested
here in a more general case.

Now, turn to the states $\left|\psi\left(\mathcal{G}\right)\right\rangle$ which are gaussian too, and will admit the same formalism. We have
\begin{equation}
\left|\psi\left(\mathcal{G}\right)\right\rangle = \underset{\mathbf{x},k}{\prod}\omega\left(\mathbf{x},k\right) \underset{\mathbf{x},k}{\prod}\mathcal{U}_{g\left(\mathbf{x},k\right)}\left(\mathbf{x},k\right) \left|A\right\rangle
\end{equation}
One can interpret now the gauging transformation $\mathcal{U}\left(\mathcal{G}\right)\equiv\underset{\mathbf{x},k}{\prod}\mathcal{U}_{g\left(\mathbf{x},k\right)}\left(\mathbf{x},k\right)$ as acting either to the right, on $\left|A\right\rangle$ (as in the main text), giving rise to the
state $\left|A\left(\mathcal{G}\right)\right\rangle = \mathcal{U}\left(\mathcal{G}\right)\left|A\right\rangle$, with the density matrix
\begin{equation}
\rho_A\left(\mathcal{G}\right) = \mathcal{U}\left(\mathcal{G}\right) \rho_A  \mathcal{U}^{\dagger}\left(\mathcal{G}\right)
\end{equation}
or the other way around, on the projection operators,
giving rise to
\begin{equation}
\rho_B\left(\mathcal{G}\right) = \mathcal{U}^{\dagger}\left(\mathcal{G}\right) \rho_B  \mathcal{U}\left(\mathcal{G}\right)
\end{equation}
Then, one obtains that the output state is
\begin{equation}
\rho\left(\mathcal{G}\right) = \text{Tr}_{V}\left( \rho_B \rho_A\left(\mathcal{G}\right)\right) = \text{Tr}_{V}\left( \rho_B\left(\mathcal{G}\right) \rho_A\right)
\end{equation}

The covariance matrix of the output gauged state will be
\begin{equation}
\begin{aligned}
\Gamma_{\text{out}}\left(\mathcal{G}\right) &= M_A\left(\mathcal{G}\right) + M_B\left(\mathcal{G}\right)\left(M_D\left(\mathcal{G}\right) - \Gamma_{\text{in}}\right)^{-1}M_B\left(\mathcal{G}\right)^T
\\&=M_A + M_B\left(M_D - \Gamma_{\text{in}}\left(\mathcal{G}\right)\right)^{-1}M_B^T
\end{aligned}
\end{equation}
where either $M$ or $\Gamma_{\text{in}}$ are transformed with respect to the gauge configuration $\mathcal{G}$. This is a very simple procedure: such transformations are mapped to orthogonal transformations on Majorana
covariance matrices \cite{Kraus2010,Zohar2015b}. Thus the covariance matrix elements may be calculated very easily using the gaussian formalism.

A very crucial quantity for our method is $\left\langle \psi\left(\mathcal{G}\right) | \psi\left(\mathcal{G}\right)\right\rangle$. This can also be calculated very simply with the gaussian formalism:
\begin{equation}
\left\langle \psi\left(\mathcal{G}\right) | \psi\left(\mathcal{G}\right)\right\rangle \propto \left\langle A\left(\mathcal{G}\right) \right| \rho_B \left| A \left(\mathcal{G}\right) \right\rangle =
\left\langle A \right| \rho_B \left(\mathcal{G}\right) \left| A  \right\rangle
\label{normcalc}
\end{equation}
where the proportion is since the $\omega$ operators are not normalized projectors, but this is irrelevant for our purposes, as we are interested eventually in
$p\left(\mathcal{G}\right) = \frac{\left\langle\psi\left(\mathcal{G}\right)|\psi\left(\mathcal{G}\right)\right\rangle}{ \int \mathcal{D}\mathcal{G}' \left\langle\psi\left(\mathcal{G}'\right)|\psi\left(\mathcal{G}'\right)\right\rangle}$. Thus one does not have to worry about the normalization in (\ref{normcalc}), and obtain simply
\begin{equation}
\left\langle \psi\left(\mathcal{G}\right) | \psi\left(\mathcal{G}\right)\right\rangle = \text{Tr}\left( \rho_B \rho_A\left(\mathcal{G}\right)\right) = \text{Tr}\left( \rho_B\left(\mathcal{G}\right) \rho_A\right)
\label{nrmtrace}
\end{equation}
which once again could be calculated using the gaussian techniques of \cite{Bravyi05}.

In the case of pure gauge theories, the norm calculation simplifies even further, as it involves no physical fermions. Thus $M=M_D$, and therefore (\ref{nrmtrace}) simply corresponds to the overlap of two gaussian states involving the same modes, e.g. $\rho_B\left(\mathcal{G}\right)$ and $\rho_A $, if we choose to act with the gauge transformation on the bonds. This has a very simple formula involving the covariance matrices \cite{Bravyi05,Mazza2012a},
\begin{equation}
 \text{Tr}\left( \rho_B\left(\mathcal{G}\right) \rho_A\right) = \sqrt{\text{det}\left(\frac{1-\Gamma_{\text{in}}\left(\mathcal{G}\right)M_D}{2}\right)}
\label{overlap}
\end{equation}
which we used in our numerical illustration, that dealt with a pure gauge theory, described next.

\section{Using the method for further observables}\label{appe}
Gauge invariant operators - which could be used as physical observables in gauge theories -  can be of several forms.
For example,
they may involve a path matrix product of $U$,$U^{\dagger}$ -
closed and traced (Wilson loop) or enclosed within the appropriate fermionic operators. Wilson loops were discussed in the main text. Here we shall comment on the other type - "meson" operators:
oriented strings of group element operators along an open path $C$, connecting fermionic operators on its edges, e.g.
\begin{equation}
\mathcal{M}\left(\mathbf{x},\mathbf{y},C\right)=\psi^{\dagger}_m\left(\mathbf{x}\right) \left(\underset{\left\{\mathbf{z},k\right\}\in C}{\prod} U\left(\mathbf{z},k\right) \right)_{mn} \psi_n\left(\mathbf{y}\right),
\end{equation}
where $C$ connects $\mathbf{x,y}$ and the $U$ matrices may get a $\dagger$ depending on the orientation, as in the Wilson Loop case.

Again, we use the fact that $\left|\mathcal{G}\right\rangle$ is an eigenstate of the gauge field part. The fermionic part may be expressed in terms of the covariance matrix, as
\begin{equation}
\left\langle \psi\left(\mathcal{G}\right)\right| \psi^{\dagger}_m \left(\mathbf{x}\right)\psi_n\left(\mathbf{y}\right)\left|\psi\left(\mathcal{G}\right)\right\rangle=
-i\mathcal{R}^{\mathcal{G}}_{nm}\left(\mathbf{y},\mathbf{x}\right)\left\langle\psi\left(\mathcal{G}\right)|\psi\left(\mathcal{G}\right)\right\rangle
\end{equation}
 We define
 \begin{equation}
F_{\mathcal{M} \left(\mathbf{x},\mathbf{y},C\right)}\left(\mathcal{G}\right)=-i\left(\underset{\left\{\mathbf{x},k\right\}\in C}{\prod}D\left(g\left(\mathbf{x},k\right)\right)\right)_{mn} \mathcal{R}^{\mathcal{G}}_{nm}\left(\mathbf{y},\mathbf{x}\right)
\end{equation}
and obtain
\begin{equation}
\left\langle \mathcal{M} \left(\mathbf{x},\mathbf{y},C\right) \right\rangle =
\int \mathcal{DG} F_{\mathcal{M} \left(\mathbf{x},\mathbf{y},C\right)}\left(\mathcal{G}\right) p\left(\mathcal{G}\right)
\end{equation}
 allowing one to use Monte-Carlo efficiently as well.

Another class of gauge invariant operators are such that are diagonal in the representation basis.
 They include, for example, local operators (on
a link) of the form
\begin{equation}
\underset{j}{\sum}f_j \Pi_j \equiv \underset{j}{\sum}f_j \left|jmn\right\rangle\left\langle jmn\right|
\end{equation}
where $f_j$ are some representation-dependent coefficients. A conventional choice for Lie groups is the casimir
operator: for example, for $U(1)$, $f_j = j^2$, and for $SU(2)$, $f_j = j\left(j+1\right)$. This operator is then understood as the electric energy on the link (since it corresponds to the square of the electric field),
which allows to write down the common \emph{Electric Hamiltonian}
\begin{equation}
H_E = \underset{\mathbf{x},k}{\sum}f_j \Pi_j\left(\mathbf{x},k\right)
\end{equation}

One can also use the method presented in the main text for such operators. Consider a gauge field operator $O\left(\mathcal{L}\right)$, which is not diagonal in terms of group element states,
acting only on a finite set of neighboring links $\mathcal{L}$ (in the electric energy case, it acts on a single link). Then,
\begin{widetext}
\begin{equation}
\left\langle\mathcal{G'}\right|O\left(\mathcal{L}\right)\left|\mathcal{G}\right\rangle =
\underset{\left\{\mathbf{x},k\right\}\notin\mathcal{L}}{\prod}\delta\left(g\left(\mathbf{x},k\right),g'\left(\mathbf{x},k\right)\right)\underset{\left\{\mathbf{x},k\right\}\in\mathcal{L}}
{\prod}f_{O\left(\mathcal{L}\right)}\left(g\left(\mathbf{x},k\right),g'\left(\mathbf{x},k\right)\right)
\end{equation}
 and if we define
\begin{equation}
F_{O\left(\mathcal{L}\right)}\left(\mathcal{G}\right)
 =
\int \mathcal{DG'}
 \left\langle\mathcal{G'}\right| O\left(\mathcal{L}\right) \left|\mathcal{G}\right\rangle \left\langle \psi\left(\mathcal{G'}\right) | \psi\left(\mathcal{G}\right)\right\rangle
/\left\langle\psi\left(\mathcal{G}\right)|\psi\left(\mathcal{G}\right)\right\rangle
\end{equation}
 we obtain a Monte-Carlo applicable form for such obervables too:
\begin{equation}
\left\langle O\left(\mathcal{L}\right) \right\rangle = \int \mathcal{DG} F_{O\left(\mathcal{L}\right)}\left(\mathcal{G}\right) p\left(\mathcal{G}\right)
\end{equation}
(as $O\left(\mathcal{L}\right)$ is local, the $\mathcal{DG'}$ integration is simple and involves only a few integration variables, and $\left\langle \psi\left(\mathcal{G'}\right) | \psi\left(\mathcal{G}\right)\right\rangle$
can be computed efficiently  as well).

Out of these building blocks one could construct the Hamiltonian of a lattice gauge theory and therefore calculate its expectation value for GGPEPS.
A particular type of a Wilson loop, $W\left(C\right)$, in case $C$ is a single plaquette (unit square), is the \emph{plaquette operator},
\begin{equation}
P\left(\mathbf{x},k_1,k_2\right)=\text{Tr}\left(
U\left(\mathbf{x},k_1\right)
U\left(\mathbf{x+e}_1,k_2\right)
U^{\dagger}\left(\mathbf{x+e}_2,k_1\right)
U^{\dagger}\left(\mathbf{x},k_2\right)\right)
\end{equation}
\end{widetext}
which allows us to write down the \emph{Magnetic Hamiltonian},
\begin{equation}
H_B = \underset{\mathbf{x},k_1<k_2}{\sum}\left(P\left(\mathbf{x},k_1,k_2\right)+P^{\dagger}\left(\mathbf{x},k_1,k_2\right)\right)
\end{equation}
Altogether, we obtain the \emph{Kogut-Susskind Hamiltonian} for lattice pure-gauge theories \cite{KogutSusskind,KogutLattice},
\begin{equation}
H_{KS} = H_E + H_B
\end{equation}

The dynamical matter terms which could be added to these are either the mass terms - local fermionic terms - for which the calculation of the expectation value does not even require Monte-Carlo integration, as it does not involve the gauge field, and gauge-matter interactions which are mesonic operators along a single link.

\section{More details on the $\mathbb{Z}_3$ illustration}\label{appf}

The $\mathbb{Z}_3$ parametrization used by us for the illustration is taken from \cite{Zohar2015b}, where it parameterized $U(1)$ gauge invariant states, with translation and rotation invariance, in two space dimensions.
However, as $\mathbb{Z}_3$ is a subgroup of $U(1)$, and since the fermionic construction imposes a truncation of the link Hilbert spaces to three dimensions, one can use the same parametrization for $\mathbb{Z}_3$ as well.

Generally, the parametrization of \cite{Zohar2015b} included dynamical fermions, but for the current work we only needed the pure gauge case, which is what we shall describe here. Therefore the state involves no physical fermions, and the $A$ operators are only used for connecting the gauge field Hilbert spaces on the links in a gauge invariant way.

There are two virtual modes on each edge (and therefore the bond dimension is $4$): $c^{j\dagger}\left(\mathbf{x},\pm k\right)$, corresponding to single copies of the representations $j=\pm1$ (no need to use $\alpha$), and $k=1,2$ - altogether eight modes. The operator $A$ is constructed using the operators
\begin{widetext}
\begin{equation}
 A = \text{exp}\left(\left(
                       \begin{array}{c}
                         c^{+\dagger}\left(\mathbf{x},-1\right) \\
                         c^{-\dagger}\left(\mathbf{x},+ 1\right)  \\
                         c^{-\dagger}\left(\mathbf{x},+2\right) \\
                         c^{+\dagger}\left(\mathbf{x},-2\right) \\
                       \end{array}
                     \right)^T
                     \left(
                       \begin{array}{cccc}
                         0 & y & z/\sqrt{2} & z/\sqrt{2} \\
                         -y & 0 & -z/\sqrt{2} & z/\sqrt{2} \\
                         -z/\sqrt{2} & z/\sqrt{2} & 0 & y \\
                         -z/\sqrt{2} & -z/\sqrt{2} & -y & 0 \\
                       \end{array}
                     \right)
                     \left(
                       \begin{array}{c}
                         c^{-\dagger}\left(\mathbf{x},-1\right) \\
                         c^{+\dagger}\left(\mathbf{x},+ 1\right)  \\
                         c^{+\dagger}\left(\mathbf{x},+2\right) \\
                         c^{-\dagger}\left(\mathbf{x},-2\right) \\
                       \end{array}
                     \right)\right)
\end{equation}
and
\begin{equation}
                     V\left(\mathbf{x},k\right) = \text{exp}\left(\left(\sigma_x\right)_{jj'}c^{j\dagger}\left(\mathbf{x},k\right)c^{j'\dagger}\left(\mathbf{x+e}_k,-k\right)\right)
 \end{equation}
 which satisfy the desired symmetry properties defined in the main text (Eqs. (\ref{transver},\ref{translink})).

One may see that the parameter $y$ connects the horizontal or the vertical virtual degrees of freedom, and therefore is responsible for creating straight flux lines. $z$, on the other hand, is
connecting horizontal degrees of freedom to vertical ones, and therefore is in charge of the corners.

For a detailed derivation of this result, the reader could refer to \cite{Zohar2015b}, where everything is explained and proven in detail. To make this easy, since we made a slight change of notations from there to be able to
generalize to higher dimensions, let us briefly comment on the notation and convention changes.

First, we changed the signs on the labels of the virtual fermions on odd sites. This does not change the operator $A$ (exchanging "positive" and "negative" operators, in the notions of \cite{Zohar2015b}), but gives a
different form to the projection operators $\omega$: in \cite{Zohar2015b}, they connected modes from opposite edges of the links, labeled by the same sign, and here the signs are opposite. This also affects the gauging procedure, as originally it was done without staggering the gauge field, and here we stagger.

Another difference is in the names of the virtual modes on a given vertex $\mathbf{x}$. This is summarized in the table below.
\begin{center}
\begin{tabular}{|c|c|c|}
  \hline
  Notation in \cite{Zohar2015b} & Current notation, $\mathbf{x}$ even & Current notation, $\mathbf{x}$ odd \\
    \hline
  \hline
  $\psi^{\dagger}\left(\mathbf{x}\right)$ & $\psi^{\dagger}\left(\mathbf{x}\right)$ & $\psi^{\dagger}\left(\mathbf{x}\right)$ \\
    \hline
  $l^{\dagger}_+\left(\mathbf{x}\right) $& $c^{+\dagger}\left(\mathbf{x},-1\right)$ & $c^{-\dagger}\left(\mathbf{x},-1\right)$ \\
    \hline
   $l^{\dagger}_-\left(\mathbf{x}\right)$ & $c^{-\dagger}\left(\mathbf{x},- 1\right)$ & $c^{+\dagger}\left(\mathbf{x},- 1\right)$ \\
     \hline
  $r^{\dagger}_+\left(\mathbf{x}\right) $& $c^{+\dagger}\left(\mathbf{x},+ 1\right)$ & $c^{-\dagger}\left(\mathbf{x},+ 1\right)$ \\
    \hline
   $r^{\dagger}_-\left(\mathbf{x}\right)$ & $c^{-\dagger}\left(\mathbf{x},+ 1\right)$ & $c^{+\dagger}\left(\mathbf{x},+ 1\right)$ \\
     \hline
  $u^{\dagger}_+\left(\mathbf{x}\right) $& $c^{+\dagger}\left(\mathbf{x},+ 2\right)$ & $c^{-\dagger}\left(\mathbf{x},+ 2\right)$ \\
    \hline
   $u^{\dagger}_-\left(\mathbf{x}\right)$ & $c^{-\dagger}\left(\mathbf{x},+ 2\right)$ & $c^{+\dagger}\left(\mathbf{x},+ 2\right)$ \\
     \hline
  $d^{\dagger}_+\left(\mathbf{x}\right) $& $c^{+\dagger}\left(\mathbf{x},- 2\right)$ & $c^{-\dagger}\left(\mathbf{x},- 2\right)$ \\
    \hline
   $d^{\dagger}_-\left(\mathbf{x}\right)$ & $c^{-\dagger}\left(\mathbf{x},- 2\right)$ & $c^{+\dagger}\left(\mathbf{x},- 2\right)$ \\
  \hline
\end{tabular}
\end{center}

 \end{widetext}

In the process of gauging, we simply put phases on the virtual fermions:
$c^{j\dagger}\left(\mathbf{x},+k\right) \rightarrow \underset{q=-1,0,1}{\sum}e^{\pm 2 \pi i (-1)^{x_1+x_2} j q /3}c^{j\dagger}\left(\mathbf{x},+k\right)\otimes\left|q\right\rangle\left\langle q \right|$
where $(-1)^{x_1+x_2}$ is due to the staggering; The $q$s are the variables which are summed
in the Monte-Carlo procedure. The relevant $U$ operators on the links are $U^{j=1} = U^{j=-1\dagger}  = \overset{q}{\underset{q=-1}{\sum}}e^{2\pi i q/3}\left|q\right\rangle\left\langle q\right|$.

As it is a pure gauge theory, the probability function could be calculated through the formula of overlap of two (virtual) fermionic gaussian states, as explained above (Eq. (\ref{overlap})).

\end{document}